\documentstyle[aps,twocolumn]{revtex}
\begin{document}
\draft
\title{Electron Cotunneling in a Semiconductor Quantum Dot}
\author{S. De Franceschi$^1$, 
S. Sasaki$^2$, J. M. Elzerman$^1$, W. G. van der Wiel$^1$, 
S. Tarucha$^{2,3}$, and L. P. Kouwenhoven$^1$}
\address{{$^1$Department of Applied Physics, DIMES, and ERATO 
Mesoscopic 
Correlation Project, Delft University of Technology,
PO Box 5046, 2600 GA Delft, The Netherlands}
\newline
{$^2$NTT Basic Research Laboratories, Atsugi-shi, Kanagawa 243-0198, 
Japan}
\newline
{$^3$ERATO Mesoscopic Correlation Project, University of Tokyo, 
Bunkyo-ku, 
Tokyo 113-0033, Japan}
}
\date{\today}
\maketitle
\begin{abstract}
We report transport measurements on a semiconductor quantum dot 
with a small number of
confined electrons. In the Coulomb blockade regime, 
conduction is 
dominated by cotunneling processes. These can be either elastic or 
inelastic, 
depending on whether they leave the dot 
in its ground state or drive it into an excited state, 
respectively. 
We are able
to discriminate between these two contributions and show that 
inelastic events can occur only if
the applied bias exceeds the lowest excitation energy. 
Implications to 
energy-level spectroscopy are discussed.   
\end{abstract}


Quantum-dot devices consist of a small electronic island connected 
by tunnel barriers to source and drain electrodes \cite{NATO}. Due to on-site 
Coulomb repulsion, the addition of an electron to the island implies an energy 
change $U=e^2/C$, where $C$ is the total capacitance of the island. 
Hence the number of confined electrons is a well-defined integer, $N$, 
that can be controlled by varying the voltage on a nearby gate electrode. 
Transport of electrons through the dot is allowed only at the transition points 
where the $N$- and ($N+1$)-states are both energetically 
accessible. Otherwise, $N$ is constant and current is strongly 
suppressed. This is known as Coulomb blockade 
\cite{NATO}. At low temperature, however, higher-order 
tunneling events can become dominant. These are commonly known as $cotunneling$ 
events since they involve the simultaneous tunneling of two or more electrons 
\cite{Averin&Nazarov}. 
Electron cotunneling has received considerable attention 
over the last decade. Initially it was recognized as 
a limitation to the accuracy of single-electron devices. 
More recently, it has acquired a broader relevance, 
especially since an increasing activity 
has been focused on quantum dots with a deliberately enhanced  
tunnel coupling to their leads. 
These systems allow the investigation of high-order 
transport processes and many-body phenomena, 
such as the Kondo effect \cite{Kondo98,Sasaki}. 
The latter can be regarded as the result of 
a coherent superposition of higher-order cotunneling events. 
Here, we will focus on the lowest order contribution 
to cotunneling.  

Previous experiments were performed with 
metallic islands 
\cite{Geerligs90,Eiles92,Hanna92} or large semiconductor dots 
\cite{Glattli91,Pasquier93,Cronenwett97}, where  the energy spectrum 
is essentially continuous 
and many levels contribute to cotunneling. Here, we 
study cotunneling through a small quantum dot where the energy 
levels are well separated, and 
where the absolute value of $N$ is precisely known.
A cotunneling event is called inelastic when it leaves 
the dot in an excited state. Otherwise it is classified as elastic. 
We identify two regimes: one consisting of elastic processes only, and one 
including both elastic and inelastic 
contributions. We note that 
the transition between these regimes can be
sharper than the characteristic 
life-time broadening of the dot states. 
In such a case, the onset of inelastic cotunneling can be exploited to 
measure the energy spectrum of a quantum dot with improved resolution. 

The stability diagram of a generic quantum dot can be obtained by 
plotting the differential 
conductance (d$I$/d$V_{sd}$) as a function of bias, $V_{sd}$, and 
gate voltage, $V_g$. Coulomb blockade occurs within the diamond-shaped 
regions in Fig. 1a.
The diamond size is proportional to the addition energy, defined as 
$E_{add}(N) \equiv \mu_{dot}(N+1)-
\mu_{dot}(N)$, where $\mu_{dot}(N)$ is the electrochemical 
potential of an $N$-electron dot. Inside the $N$-electron diamond, 
$\mu_{dot}(N) < \mu_L, \mu_R < \mu_{dot}(N+1)$, with $\mu_L$, $\mu_R$ the 
Fermi energies of the leads. 
The diamond edges correspond to level alignment: $\mu_{dot}(N)$ = $\mu_L$ or 
$\mu_R$ (see angled solid lines). This alignment determines the onset for 
first-order tunneling via the ground state of the dot, leading to a peak in 
d$I$/d$V_{sd}(V_{sd})$. The onset for first-order tunneling via the first 
excited state occurs at a somewhat higher bias (see dot-dashed lines 
in Fig. 1a, and the  
corresponding diagrams in Fig. 1b and 1e). These first-order 
processes have been exploited as a spectroscopic tool on the discrete energy 
spectrum of dots \cite{NATO}.
	
Here, we are interested in second-order tunneling of charge, which becomes 
more apparent when the tunnel coupling between the dot and the leads 
is enhanced. 
We neglect 
contributions from spin that could give rise to the Kondo effect. Elastic 
cotunneling is the dominant off-resonance process at low bias. It gives 
rise to current inside the Coulomb diamond (light-grey region in Fig. 1a). The 
corresponding two-electron process (Fig. 1c) transfers one electron from the 
left to the right lead, thereby leaving the dot in the ground state. 

For $e|V_{sd}| \ge \Delta(N)$, where $\Delta(N)$ is the lowest on-site 
excitation 
energy for a constant $N$ \cite{note1}, similar two-electron 
processes can occur which drive the dot into an excited state. For instance, an 
electron can leave the dot from the ground state to the lowest Fermi sea,
while another electron from the highest Fermi sea tunnels into 
the excited 
state (see Fig. 1d). Although this type of  process is called inelastic 
\cite{Averin&Nazarov}, the total electron energy is 
conserved. The on-site excitation is created at the expense of 
the energy drop $eV_{sd}$. To first approximation, the onset of inelastic 
cotunneling yields a step in d$I$/d$V_{sd} (V_{sd})$ \cite{Funabashi99}. This 
step occurs when $e|V_{sd}| = \Delta(N)$, which is not or only weakly affected 
by $V_g$ (see also Fig. 1c in Ref. \cite{Schmid00}). 
As a result, inelastic cotunneling turns on along the 
vertical (dotted) lines in Fig. 1a. At the edge of the Coulomb diamond the 
condition for the onset of inelastic cotunneling connects to that for 
the onset of first-order tunneling via an excited state (dot-dashed lines).

Our device has the external shape of a 0.5-$\mu$m-high pillar with a 
$0.6 \times 0.45$ $\mu$m$^2$ rectangular base (inset to Fig. 2). 
It is fabricated from an 
undoped AlGaAs(7 nm)/InGaAs(12 nm)/AlGaAs(7 nm) double barrier heterostructure, 
sandwiched between n-doped GaAs source and drain electrodes \cite{Sasaki}. 
The quantum dot is 
formed within the InGaAs layer. The lateral confinement potential is close to 
that of an ellipse \cite{Austing99}. Its strength is tuned by a negative 
voltage, $V_g$, applied to a metal gate surrounding the pillar. A dc bias 
voltage, $V_{sd}$, applied between source and drain, drives current 
vertically 
through the pillar. In addition, we apply a small bias modulation with rms 
amplitude $V_{ac} = 3$ $\mu$V at 17.7 Hz for lock-in detection. 
Measurements are 
carried out in a dilution refrigerator with a base temperature of 15 mK. 
We find 
an effective electron temperature $T_e= 25 \pm 5$  mK, due to residual 
electrical noise.   

Figure 2 shows d$I$/d$V_{sd}$ in grey-scale versus ($V_{sd}$,$V_g$) at 15 mK. 
Diamond-shaped regions of low conductivity (light grey) identify the Coulomb 
blockade regimes for $N$ = 1 to 4. The diamonds are delimited by dark narrow 
lines (d$I$/d$V_{sd} \sim e^2/h$) 
corresponding to the onset of first-order tunneling. 
For $N=1$, as well as for $N=3$, sub-gap transport is dominated by elastic 
cotunneling with no evidence for inelastic cotunneling.
The differential conductance is uniformly low inside the Coulomb 
diamond. (Slight modulations are seen due to a weak charging effect in the GaAs 
pillar above the dot \cite{foot1}.) 
This is different for $N=2$, where the onset of 
inelastic cotunneling is clearly 
observed. As argued before, this onset follows (dotted) lines, nearly parallel 
to the $V_{g}$ axis \cite{foot3}. 
At the diamond edges they connect to (dot-dashed) 
lines where 
first-order tunneling via an excited state sets in. 
Similar considerations apply to $N=4$.

The different behavior observed for $N=even$ and $N=odd$ 
stems from the fact 
that inelastic cotunneling occurs only if $E_{add}(N) > \Delta(N)$,
as apparent from Fig. 1a. In the case of non-interacting electrons, 
$E_{add}(N) = U(N)$ for $N=odd$ and $E_{add}(N) = U(N) + \Delta(N)$ 
for $N=even$, where $U(N)$ is the charging energy for $N$ electrons;
energy levels are spin degenerate
 and consecutively filled with pairs of electrons. 
This is a reasonable picture if the level spacing exceeds 
the exchange interaction energy \cite{Tarucha00}. In our small 
quantum dot the first three levels are indeed widely spaced as 
already discussed in Ref. \cite{Sasaki}. For low $N$, $\Delta(N)$ 
exceeds not only the exchange energy but also $U(N)$. 
This implies that for $N=odd$, $\Delta(N)$ 
lies outside the Coulomb diamond (i.e. $\Delta(N) > E_{add}(N)$) 
and thus inelastic cotunneling is not 
observed. 
(Note that cotunneling between spin-degenerate states is an 
elastic process as initial and final state have the same energy.)
For $N=even$, $\Delta(N)$ is always smaller than $E_{add}(N)$ 
and inelastic cotunneling can be observed. 

We now discuss the difference in life-time broadening between first- and 
higher-order tunneling. At the onset of first-order tunneling a certain 
level is 
aligned to one of the Fermi energies. In this case, an electron can escape from 
the dot, which leads to a finite life-time broadening of the observed resonance 
by an amount $\hbar \Gamma$. Here, $\Gamma = \Gamma_L + \Gamma_R$, where 
$\Gamma_L$ and $\Gamma_R$ are the tunneling rates through the left and 
the right barrier, respectively. 
(Note that these rates are independent of $V_{sd}$, since 
our bias window ($\sim$meV) 
is much smaller than the height of the 
AlGaAs tunnel barriers ($\approx$50 meV). 

The onset of inelastic cotunneling is also characterized by a 
width. In the zero-temperature limit, this is determined by the life-time 
broadening of the excited state. 
Two types of situations can occur. First, the excited state can be 
between $\mu_L$ and $\mu_R$ (see right inset to Fig. 3) so that 
inelastic cotunneling can 
be followed by first-order tunneling. 
Such a decay event leads to a life-time broadening of at least 
$\hbar \Gamma_R \approx \hbar \Gamma/2$. Second, the ground and 
excited state are 
both well below $\mu_L$ and $\mu_R$, implying 
that only higher-order tunneling is 
allowed (see right inset to Fig. 4). Now, decay from 
the excited state can only rely on cotunneling. 
For this higher-order perturbation, 
the corresponding rate, $\Gamma_{co}$, is much smaller than $\Gamma$, 
leading to a reduced life-time broadening.   
To illustrate these arguments, 
we select different d$I$/d$V_{sd}-vs-V_{sd}$ traces and analyse their shape 
in detail. 

Figure 3 shows two traces for $N=2$, taken at 15 mK for gate voltages at the 
horizontal lines in the left inset. The dashed trace has several peaks. The two 
inner ones, at $|V_{sd}| \approx 1.1$ mV, correspond 
to first-order tunneling of the 3rd electron via the 
3-electron ground state; i.e. $\mu_{dot}(3)$ = $\mu_L$ or $\mu_R$. 
The right (left) peak has a full width 
at half maximum (FWHM) of $\approx$200 ($\approx$400) $\mu$V. 
This is somewhat larger than the 
width, $\hbar \Gamma/e \simeq 150$ $\mu$V, measured in the zero-bias limit. 
Indeed a finite bias allows 
non-energy-conserving tunneling events leading to additional broadening. 
The most likely source 
for energy relaxation is acoustic-phonon emission \cite{Fujisawa}. The 
following pair of 
peaks, at  $|V_{sd}| \approx 2$ mV, 
corresponds to the onset of first-order tunneling via the first 
excited state for $N=2$ (see Fig. 1b). 
Because of the larger bias voltage, 
these peaks are visibly broader than the 
inner ones. 
Additional peak structures occur near the 
edges of the bias window. 
The origin of these peaks can not be precisely identified. 

The solid trace contains structure from both first- and second-order tunneling. 
The peaks labeled by open squares arise from first-order tunneling at 
the edges of the Coulomb diamond (see left inset).  
Steps, labeled by open circles, identify the 
onset of inelastic cotunneling and correspond to the 
open circles in the left inset.  
Their different heights are probably due to a left-right asymmetry 
in the tunnel coupling to the leads. 
Their $V_{sd}$-position, which is 
symmetric around zero bias, provides a direct measure of $\Delta(2)$. The 
width of these steps is $\approx$150 $\mu$V \cite{foot2}.
Since $\Delta(2) \approx U(2)$, the first excited state lies unavoidably within 
the bias window  when $|V_{sd}| = \Delta(2)/e$ and hence is allowed 
to decay into 
the lowest-energy lead (see the right inset to Fig. 3). As argued above, this 
situation leads to a step-width exceeding $\hbar \Gamma_R/e$, consistent 
with our finding.  
Another structure occurs at $V_{sd} \approx 2.6$ mV and is probably due to 
the onset of inelastic cotunneling via the second excited state for $N=2$. The 
corresponding line in the stability diagram is hardly visible due to its 
vicinity to the diamond edge. 

To study inelastic cotunneling when both ground and excited state lie
well below the Fermi energies of the leads (Fig. 1d) we need $\Delta(N)
\ll E_{add}(N)$. To this aim we move to 
$N=6$, since $\Delta(6)$ can be effectively tuned 
by a magnetic field applied along the vertical axis. 
We tune the field to 0.35 T, such that $\Delta(6) \approx 0.1$ meV, 
i.e. several times 
smaller than $E_{add}(6)$. From a previous study we know that 
the ground state is a
spin singlet, and the first excited state is a spin triplet \cite{Sasaki}.
The d$I$/d$V_{sd}-vs-V_{sd}$ traces shown 
in Fig. 4 are taken at two different temperatures, but for the same $V_g$, at 
the horizontal line in the left inset. 
The solid trace (15 mK) shows a broad minimum around $V_{sd}=0$, 
where transport is dominated by elastic 
cotunneling via the ground state (see also the light-grey region in the left 
inset). 
The differential conductance increases rapidly at the onset of 
inelastic cotunneling with a step-width of 
$\approx$20 $\mu$V, i.e. much smaller  $\hbar \Gamma/e$. 
This reduced width stems from the fact that the  
excited state can not decay directly into the lower energy lead (see right 
inset). 
The corresponding life-time broadening, $\hbar \Gamma_{co}$, 
can be estimated from the 
cotunneling current, $I_{co}$, at $V_{sd}=\Delta(6)/e$.
We find $\hbar \Gamma_{co} = \hbar I_{co}/e \approx (\hbar/e) 
\int_0^{\Delta(6)/e}$ d$I$/d$V_{sd} (V_{sd})$ 
d$V_{sd} \simeq 10$ $\mu$eV, consistent with the observed step-width.
At $T_e = 25$ mK  
the thermal broadening of the Fermi distribution 
functions leads to a step-width of 
$5.44 k_B T_e/e \simeq 12 $ $\mu$eV \cite{Wolf}. 
Hence life-time broadening has been reduced here to the thermal limit.

The cotunneling onset in Fig. 4 shows a peak structure at low temperature 
(solid trace) in addition to the expected step structure. 
This is likely due to Kondo correlations, 
as discussed in Ref. \cite{Sasaki}. 
On increasing temperature to 200 mK 
these Kondo correlations are suppressed such that 
only lowest order cotunneling contributes. This recovers the   
step structure (dashed trace). 

We thank Yu. V. Nazarov, M. R. Wegewijs, M. Eto, K. Maijala, 
and J. E. Mooij  for discussions. 
We acknowledge financial support from the Specially Promoted 
Research, 
Grant-in-Aid for Scientific Research, from the Ministry of 
Education, 
Science and Culture in Japan, from the Dutch Organisation for 
Fundamental Research on Matter (FOM), from the NEDO joint research 
program (NTDP-98), and from the EU via a TMR network.

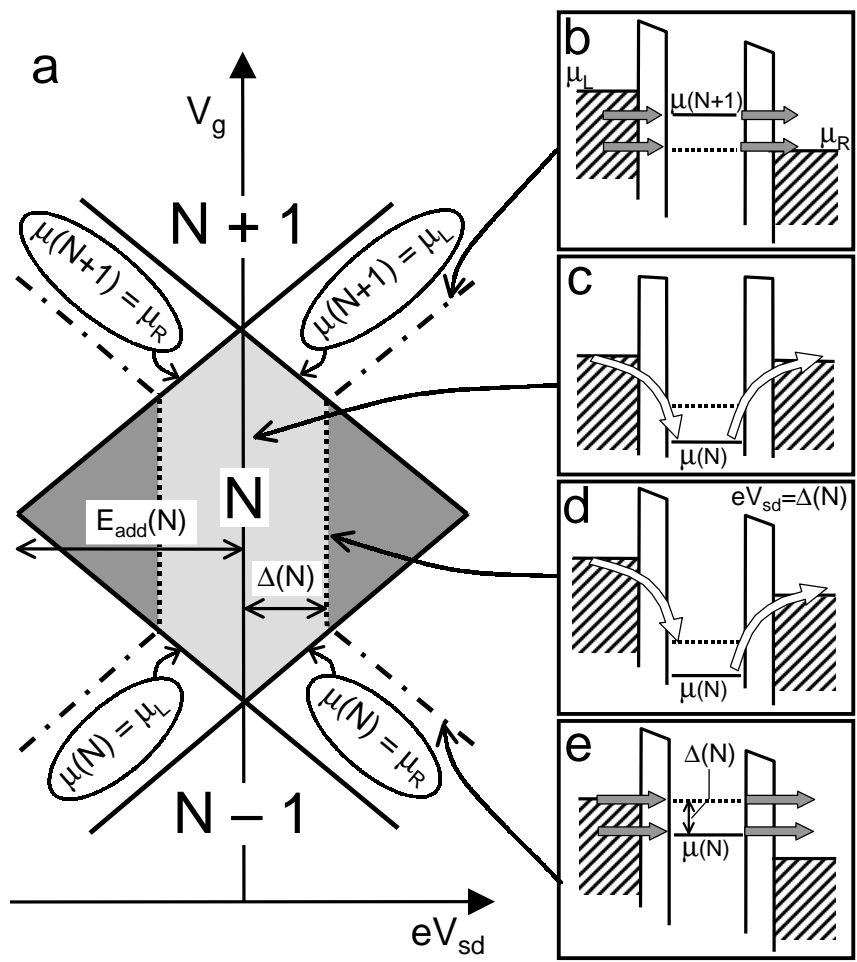
\begin{figure}
\caption{
(a) Stability diagram in the plane of ($V_{sd}$, $V_g$). 
Angled lines correspond 
to $alignment$ of a dot-state with the Fermi energy of the leads. In this case, 
first-order tunneling sets in, or is increased, as illustrated in (b) and (e). 
In the light-grey area in (a), conduction is due to elastic cotunneling 
via virtual events as shown in (c). 
For $e V_{sd} \ge \Delta(N)$, inelastic processes, 
illustrated in (d), increase the 
cotunneling current (dark-grey 
areas). $\Delta(N)$ is the energy spacing between the ground 
state and the first excited state, which in (b)-(e) are represented 
by solid and dotted lines, respectively. 
}
\label{F1}
\end{figure}

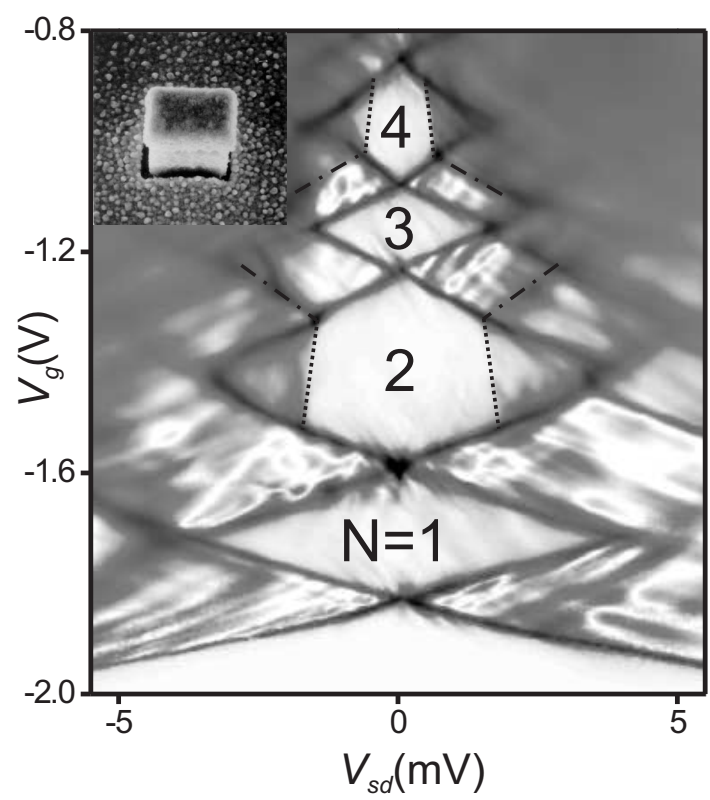
\begin{figure}
\caption{
Measured stability diagram of our quantum dot at 15 mK and zero magnetic field. 
d$I$/d$V_{sd}$ is plotted in grey scale as a function of ($V_{sd}$, $V_g$). 
Dotted lines have been superimposed to highlight the onset of inelastic 
cotunneling. The dot-dashed lines indicate the onset of first-order tunneling 
via an excited state. Inset: scanning electron micrograph of the device. 
}
\label{F2}
\end{figure}

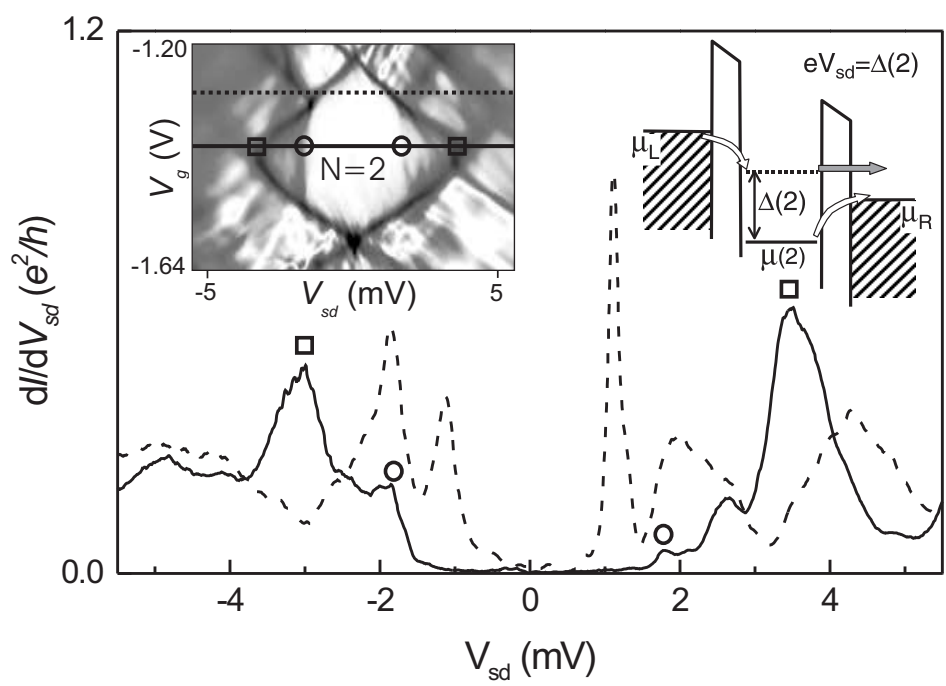
\begin{figure}
\caption{
Differential conductance as a function of bias for $V_g=-
1.40$ V (solid line) and 
$V_g=-1.30$ V (dotted line) at 15 mK. These traces are extracted from the 
stability diagram shown in the left inset. 
The horizontal lines indicate the corresponding $V_g$ values. 
The right inset shows the 
qualitative energy diagram corresponding to the onset of inelastic cotunneling 
for $N=2$. The horizontal arrow represents the possibility for an electron in 
the excited state to decay directly into the right lead by first-order 
tunneling. 
}
\label{F3}
\end{figure}

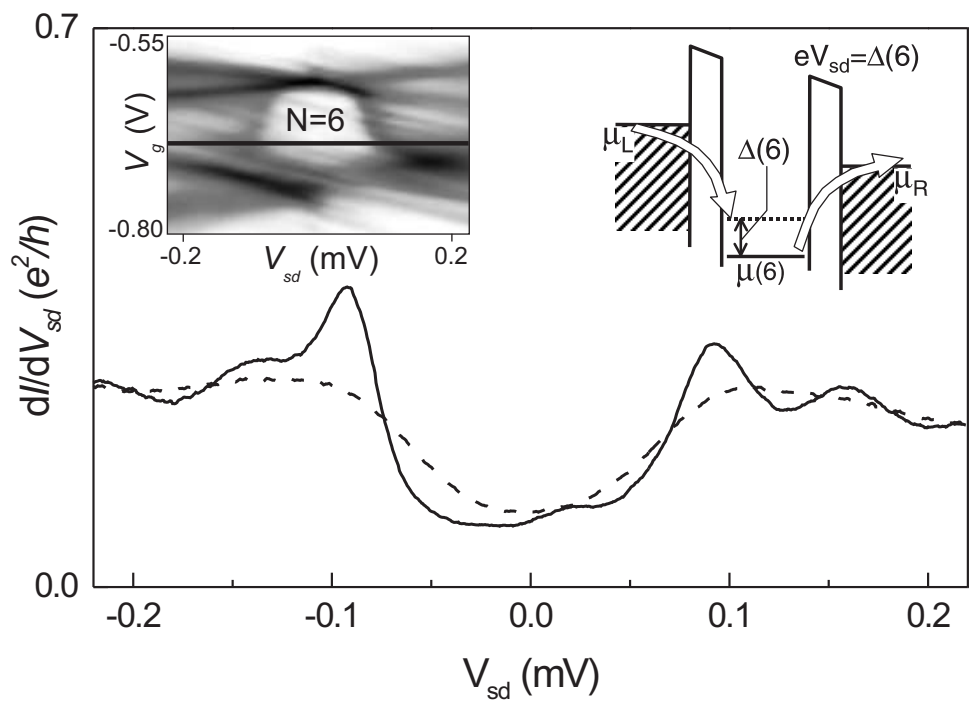
\begin{figure}
\caption{
Differential conductance as a function of bias 
at $V_g=-0.685$ V. (Note that the bias window is much 
smaller than in Fig. 3.)
The solid (dashed) line is taken at 15 mK (200 mK). 
Left inset: stability diagram at 
15 mK, around the 6-electron Coulomb diamond. 
The horizontal line is at $V_g=-0.685$ V.
Right inset: qualitative energy diagram corresponding
to the onset of inelastic cotunneling for $N=6$.
}
\label{F4}
\end{figure}


\begin{references}

\bibitem{NATO}
L. P. Kouwenhoven, C. M. Marcus, P. L. McEuen, S.
Tarucha, R. M. Westervelt, and N. S. Wingreen, 
in {\it Mesoscopic Electron Transport}, edited by L.L. Sohn,
L. P. Kouwenhoven, and G. Sch\"{o}n, (Kluwer, Series E 345, 1997), 
p.
105-214.    

\bibitem{Averin&Nazarov}
D. V. Averin and Yu. V. Nazarov, in {\it Single Charge Tunneling - 
Coulomb Blockade Phenomena in Nanostructures}, edited by H. 
Grabert and M. H. Devoret 
(Plenum Press and NATO Scientific Affairs Division, New York, 
1992), p. 217.   

\bibitem{Kondo98}
D. Goldhaber-Gordon {\it et al.}, Nature {\bf 391}, 156, (1998); 
Cronenwett {\it et al.}, Science {\bf 281}, 540 (1998); 
Schmid {\it et al.}, Physica B {\bf 256-258}, 182 (1998). 

\bibitem{Sasaki} 
S. Sasaki {\it et al.}, Nature {\bf 405}, 764 (2000). 

\bibitem{Geerligs90}
L. J. Geerligs, D. V. Averin, and J. E. Mooij, Phys. Rev. Lett. 
{\bf 65},
3037 (1990).

\bibitem{Eiles92} 
T. M. Eiles {\it et al.}, Phys. Rev. Lett. {\bf 69}, 148 (1992). 

\bibitem{Hanna92}
A. E. Hanna, M. T. Tuominen, and M. Tinkham, Phys. Rev. Lett. {\bf 
68}, 
3228 (1992). 

\bibitem{Glattli91}
D. C. Glattli {\it et al.}, Z. Phys. B {\bf 85}, 375 (1991).

\bibitem{Pasquier93}
C. Pasquier {\it et al.}, Phys. Rev. Lett. {\bf 70}, 69 (1993).

\bibitem{Cronenwett97}
S. M. Cronenwett {\it et al.}, Phys. Rev. Lett. {\bf 79}, 2312 
(1997).

\bibitem{note1} 
In our definition $\Delta(N)$ is strictly positive.
It reduces  
to the single-particle level spacing for non-interacting electrons
(for instance, in the case of $N=1$ and $N=2$, $\Delta(N)$ 
is the spacing between the first two single-particle 
levels).    

\bibitem{Funabashi99}
Y. Funabashi {\it et al.}, Jpn. J. Appl. Phys. {\bf 38}, 388 
(1999). 

\bibitem{Schmid00}
J. Schmid {\it et al.}, Phys. Rev. Lett. {\bf 84}, 5824 (2000).

\bibitem{Austing99}
D. G. Austing, {\it et al.}, Phys. Rev. B {\bf 60}, 11514 (1999).

\bibitem{foot1}
The top contact is obtained by deposition of Au/Ge and 
annealing at 400 $^\circ$C for 30 s. This thermal treatment is gentle 
enough 
to prevent the formation of defects near the dot, but does not 
allow 
the complete suppression of the native Schottky barrier. 
The residual barrier leads to electronic confinement and 
corresponding charging effects in the GaAs pillar.

\bibitem{foot3}
$V_g$ affects not only the bottom but also the shape of the 
confining potential.   
As a result, 
the level spacing (and hence $\Delta(N)$) depends weakly on $V_g$,
leading to a non-zero angle between the dotted lines and 
the $V_g$ axis.

\bibitem{Tarucha00}
S. Tarucha {\it et al.}, Phys. Rev. Lett. {\bf 84}, 2485 (2000).


\bibitem{Fujisawa} 
T. Fujisawa {\it et al.}, Science {\bf 282}, 932 (1998). 

\bibitem{foot2} The step-width is estimated by taking the 
full width at half maximum 
of the corresponding peak (or dip) in d$^2 I$/d$V_{sd}^2(V_{sd})$.

\bibitem{Wolf} 
E. L. Wolf, {\it Principles of Electron Tunneling Spectroscopy}, 
(Oxford,
New York, 1985) p. 438.

\end{references}
\end{document}